\documentclass[fleqn,twoside]{article}
\usepackage{amsmath}
\usepackage{espcrc2}
\usepackage{graphicx}
\usepackage{bbm}

\newcommand{\bb}[1]{\mathbbm{#1}}

\newcommand{\Z}{{\bb{Z}}}

\newcommand{\DD}{{\mathrm{D}}}
\newcommand{\ee}{{\mathrm{e}}}
\newcommand{\ii}{{\mathrm{i}}}

\newcommand{\SO}{{\mathrm{SO}}}
\newcommand{\SU}{{\mathrm{SU}}}
\newcommand{\Tr}{\operatorname{Tr}}
\newcommand{\sgn}{\operatorname{sgn}}
\newcommand{\Centre}{\operatorname{Centre}}
\newcommand{\F}{{\text{F}}}
\newcommand{\A}{{\text{A}}}

\newcommand{\tbc}{{\text{tbc}}}
\newcommand{\pbc}{{\text{pbc}}}

\title{Vortex free energies in SO(3) and SU(2) lattice gauge theory
\thanks{Talk presented by O. Jahn at Lattice 2002.}}

\author{Ph.\ de Forcrand%
  \address[ETH]{Institute for Theoretical Physics, 
    ETH Z\"{u}rich, CH-8093 Z\"{u}rich, Switzerland}%
  \address{CERN, Theory Division, CH-1211 Gen\`{e}ve 23, Switzerland}%
  \ and O. Jahn%
  \addressmark[ETH]%
  }

\begin{document}

\begin{abstract}
  Lattice gauge theories with gauge groups SO(3) and SU(2) are
  compared.  The free energy of electric twist, an order parameter for
  the confinement-deconfinement transition which does not rely on
  centre-symmetry breaking, is measured in both theories.  The results
  are used to calibrate the scale in SO(3).\vspace{-1ex}
\end{abstract}

\maketitle
\mathindent=0pt

\section{What makes SO(3) interesting}

The motivation for a comparison of SO(3) and SU(2) lattice gauge theory
is to clarify an apparent paradox.  On one hand, both groups have the
same local structure, $\SO(3)\!=\!\SU(2)/\Z_2$, so the naive continuum
limit of both theories is the same.  Universality of the continuum
fixed point suggests that this should be true also non-perturbatively.
On the other hand, the deconfinement transition is usually associated
with the breakdown of the centre symmetry, which does not exist in
SO(3).  If SO(3) has a deconfinement transition, its characterisation
must be different; and there should also be another order parameter.

Further doubt about universality comes from the observation~\cite{DG}
that the weak coupling phase of SO(3) LGT features 2 metastable states:
in addition to the state with positive adjoint Polyakov loop expected
in a deconfined phase, there is a state with negative adjoint Polyakov
loop.  This state has not found a satisfactory explanation so far.

\section{Centre vortices}

Even though the centre of SO(3) is trivial, centre vortices exist both
in SU(2) and SO(3).  A centre vortex is a 2-dimensional configuration
which tends to a pure gauge at infinity.  In SU(2), the corresponding
gauge function $g$ changes its sign as one goes around the vortex.  In
SO(3), the sign of the group elements is discarded, so $g(\varphi)$
becomes single-valued, but now it is a non-trivial element of
$\pi_1[\SO(3)]$.  For general gauge groups $G$, the relevant quantity
is $\pi_1[G/\Centre(G)]$.

A convenient way to study centre vortices in SU(2) are twisted
boundary conditions on a torus.  These are imposed by introducing a
mismatch in the transition functions along 2 directions, $\Omega_\mu
\Omega_\nu = - \Omega_\nu \Omega_\mu$.  
These boundary conditions change the number of centre vortices through
the plane with twist from even to odd.  Accordingly, one can define the
free energy of an (additional) centre vortex as the ratio of partition
functions with twisted and periodic boundary conditions
\begin{equation}
  \label{eq:F}
  \ee^{-\beta F_{\text{CV}}} = Z^\tbc/Z^\pbc \;.
\end{equation}
For electric twist, this ratio is an order parameter of the
deconfinement transition: for large volumes, it tends to 1 in the
confined phase while it is exponentially suppressed in the deconfined
phase \cite{LAT01}.

\section{Twist in SO(3)}

In SO(3), twisted boundary conditions are void because the sign of the
transition functions is discarded; twist becomes an ordinary
topological quantum number.  In order to understand this in more
detail, we invoke the formulation of Wilson SU(2) in terms of
SO(3) and $\Z_2$ variables \cite{MP},
\begin{multline}
  \label{eq:MPKT}
    Z^{\text{Wilson}}_{\SU(2)} = \sum_{\alpha_p=\pm1} \int_{\SU(2)}
    \!\!\! \DD U \, \ee^{\frac12\beta\sum_p \alpha_p \Tr_\F U_p} 
    \\[-1.5ex] \times
    \prod_{c\in\text{cubes}} \delta\biggl( \prod_{p\in\partial c}^6 \alpha_p - 1
    \biggr) \;.
\end{multline}
The constraint forbids monopoles in the plaquette field $\alpha$. 
Without the constraint, one obtains the Villain partition function of
SO(3) LGT,
\begin{equation}
  \label{eq:Villain}
    Z^{\text{Villain}}_{\SO(3)} = \!\!\sum_{\alpha_p=\pm1}
    \int_{\SU(2)} \!\!\!\DD U \,
    \ee^{\frac12\beta\sum_p\alpha_p\Tr_\F U_p} .
\end{equation}
The sum over $\alpha_p$ makes the action independent of the sign of
each link matrix $U$, so (\ref{eq:Villain}) is really a redundant
formulation of an SO(3) theory.

So far the argument holds for an infinite volume.  On a 4-torus, 6
additional global constraints are required \cite{Tomboulis},
namely
\begin{equation}
\label{eq:global-constraint}
  \prod_{p\in\mu\nu\text{-plane}} \alpha_p \stackrel{!}{=} +1
  \qquad\text{(periodic SU(2)).}
\end{equation}
Here, the product is over all plaquettes in a fixed 2-dimensional
section in $\mu\nu$-direction of the torus.  Equation
(\ref{eq:global-constraint}) ensures periodic boundary conditions in
SU(2).  If the product is negative for some $\mu,\nu$, one obtains
twisted boundary conditions in that direction.  For then one $U_p$ in
each $\mu\nu$-plane enters the action with a negative sign.

\begin{figure}[thb]
\centering
\mbox{\includegraphics[width=\columnwidth,trim=-15 5 10 8]{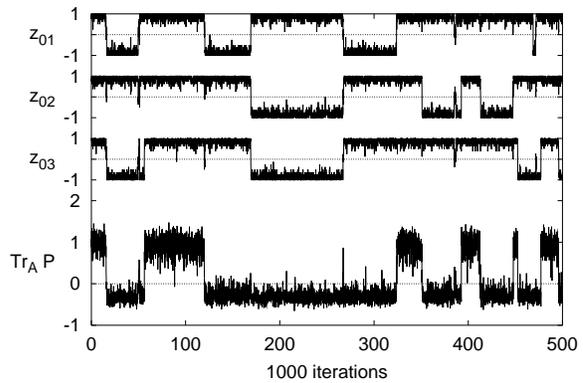}}%
\vspace*{-4ex}
\caption{Monte Carlo history of the 3 electric twist variables (top) and 
the adjoint Polyakov loop (bottom) ($4^4$ lattice, $\beta=4.5$).\vspace*{-1ex}}
\label{fig:MChist}
\end{figure}

The above relation suggests the following definition of a ``twist''
observable in the SO(3) theory (with monopoles):
\begin{equation}
  \label{eq:twist-so3}
  z_{\mu\nu} \equiv \frac{1}{L_\rho L_\sigma} \sum_{\mu\nu\text{-planes}}
  \,\,\prod_{\text{plane}} \sgn \Tr_\F U_p \;.
\end{equation}
At weak coupling, $\sgn \Tr_\F U_p$ and $\alpha_p$ are strongly
correlated; $z_{\mu\nu}$ is a genuine SO(3) observable because each
link appears twice in the product, so its sign drops out.
The average over parallel planes has been introduced because, in the
presence of monopoles, the twist can vary between planes.  So $z$ in
general takes fractional values.  It turns out that it is always close
to $\pm1$ above the bulk phase transition at $\beta\approx 4.45$ below
which monopoles condense.  This is illustrated in Fig.\ 
\ref{fig:MChist} by a Monte Carlo history obtained at $\beta=4.5$.

We also find a simple explanation for the mysterious phase with
negative adjoint Polyakov loop as a state with electric twist.  A
classical ground state in the sector with $z_{01}=-1$, say, can be
obtained by setting all links to 1, except $U_1(1,y,z,t)=\ii\sigma_1$
and $U_4(x,y,z,1)=\ii\sigma_2$ \cite{twisteater}.  The Polyakov loop
of this configuration is $\Tr_\F P=0$ in the fundamental
representation and $\Tr_\A P\equiv (\Tr_\F P)^2-1=-1$ in the adjoint
representation.  So we expect a negative $\Tr_\A P$ in sectors with
electric twist.  Figure \ref{fig:MChist} shows that this is indeed the
case.

\section{Vortex free energies}

We use the observable (\ref{eq:twist-so3}) to measure the free energy
of electric twists in the Villain SO(3) theory.  The technical
difficulty here is that there are high barriers between the different
twist sectors, so it is very difficult to maintain ergodicity.
The density of states as a function of the 3 twist variables $z_{0 i}$
extends over 12 orders of magnitude \cite{StaraLesna}.
The remedy is a multicanonical algorithm where
the barriers are removed by a bias, which is corrected by reweighting
the observables.  The bias depends on 3 variables ($z_{0 i}$)
and is represented by a 3-dimensional table determined iteratively.

\begin{figure}[thb]
\centering
\vspace{7pt}
\mbox{\includegraphics[width=\columnwidth,trim=8 0 54 0]{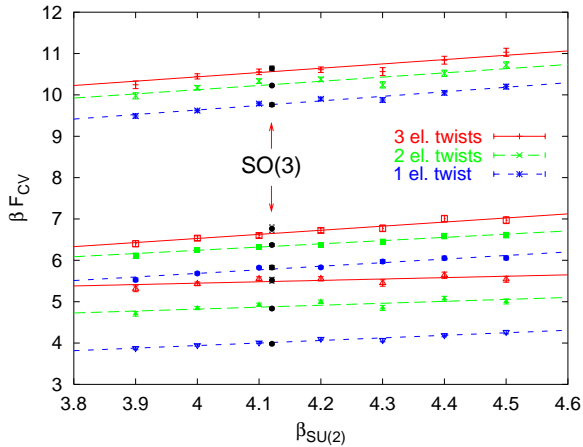}}%
\vspace{-4ex}
\caption{Comparison of the vortex (electric twist) free energies between $SO(3)$
  at $\beta=4.5$ and $SU(2)$ at various $\beta$.
Lattice sizes are $8^3$, $6^3$ and $4^3\times4$ from the top (3 lines each).%
\vspace*{-1ex}}
\label{fig:match}
\end{figure}

In Fig.\ \ref{fig:match}, the free energies of 1, 2 and 3 electric
twists (an additional vortex winding around 1, 2 or 3 directions)
obtained at $\beta_{\SO(3)}=4.5$ on $4^3$, $6^3$ and $8^3\times4$
lattices are compared with results for SU(2) at various
$\beta_{\SU(2)}$.  The latter were obtained with the method of
\cite{PRL}.  The SO(3) data can be reproduced by SU(2) with
$\beta_{\SU(2)}=4.12(3)$.  The good quality of the joint fit
($\chi^2/\text{dof}=1.35$) lends support to the hypothesis of
universality of the continuum limit.

Asymptotic scaling suggests that the inverse lattice spacing at this
coupling is about 200 GeV.  As the bulk transition prohibits coarser
lattices, we conclude that lattices larger than about $700^4$ are
needed to simulate confined (Villain) SO(3).  We would like to
emphasise that the scale of 200 GeV is in no way related with continuum
physics.  It is just due to the bulk phase transition of the SO(3)
lattice gauge theory beyond which lattice artifacts dominate, and gives
a lower bound on the possible cutoffs one can use.  This value can be
shifted by suppressing (or enhancing) lattice artifacts \cite{MMP}.

The fact that the couplings in SU(2) and SO(3) are not very different
is not a coincidence: since the actions of the two theories differ
only by terms exponentially small in the coupling, the perturbative
scale parameters $\Lambda_{\text{L}}$ are the same for both.  The
difference is thus of purely non-perturbative origin.

\section{Conclusions}

To conclude, we have recalled that SO(3) features twist sectors like
SU(2); but unlike SU(2) they are summed over within periodic boundary
conditions rather than imposed by the boundary conditions.  The vortex
free energy can serve as an order parameter in both theories, so the
centre symmetry -- which does not exist in SO(3) -- is not needed.
The free energies of spatial centre vortices are very similar in SO(3)
and SU(2) provided the bare couplings are adjusted.  This supports
universality of the continuum limit.  However, one has to keep in mind
that the systems studied here are very small.  Because of a bulk phase
transition, one needs lattices larger than $700^4$ to simulate the
confined phase.

The vortex free energy can be defined whenever
$\pi_1[G/\Centre(G)]\ne1$, independent of whether there is a centre or
not, i.e.\ for all simple Lie groups except G$_2$, F$_4$ and E$_8$.
The absence of a vortex order parameter in the latter prompts
speculations about the nature of the confinement/deconfinement phase
transition.  The study of G$_2$ proposed in \cite{G2} will be
interesting in this connection.

\newcommand{\jour}[1]{#1}
\newcommand{\volume}[1]{#1}


\begin{thebibliography}{99}

\bibitem{DG}
S. Cheluvaraja and H.\,S. Sharathchandra,
%Finite temperature properties of mixed action lattice gauge theory,
hep-lat/9611001;
S. Datta and R.\,V. Gavai,
%Phase transitions in SO(3) lattice gauge theory,
Phys.\ Rev.\ D 57 (1998) 6618.
%hep-lat/9708026.
%%CITATION = HEP-LAT 9708026;%%

\bibitem{LAT01}
Ph.~de Forcrand and L. von Smekal,
%'t Hooft loops and consistent order parameters for confinement,
Nucl.\ Phys.\ Proc.\ Suppl.\ 106 (2002) 619;
%[arXiv:hep-lat/0110135].
%%CITATION = HEP-LAT 0110135;%%
%
%\bibitem{LvS}
%Ph.~de Forcrand and L. von Smekal,
%``'t Hooft loops, electric flux sectors and confinement in SU(2)  Yang-Mills theory,''
Phys.\ Rev.\ D 66 (2002) 011504.
%[arXiv:hep-lat/0107018].
%%CITATION = HEP-LAT 0107018;%%

\bibitem{MP}
G. Mack and V.\,B. Petkova,
%Z2 Monopoles In The Standard SU(2) Lattice Gauge Theory Model,
Z.\ Phys.\ C 12 (1982) 177;
%%CITATION = ZEPYA,C12,177;%%
%
%\bibitem{Tomboulis}
E. Tomboulis,
%The 'T Hooft Loop In SU(2) Lattice Gauge Theories,
Phys.\ Rev.\ D 23 (1981) 2371.
%%CITATION = PHRVA,D23,2371;%%

\bibitem{Tomboulis}
E.\,T.~Tomboulis,
%``Confinement Via Dynamical Monopoles,''
Phys.\ Lett.\ B 303 (1993) 103;
%%CITATION = PHLTA,B303,103;%%
%
%\bibitem{KT-JMP}
T.\,G. Kov\'acs and E.\,T. Tomboulis,
%``Absence of confinement in the absence of vortices,''
J.\ Math.\ Phys.\ 40 (1999) 4677.
%[arXiv:hep-lat/9806030].
%%CITATION = HEP-LAT 9806030;%%

\bibitem{twisteater}
J. Ambj\o{}rn and H. Flyvbjerg,
%'t Hooft's nonabelian magnetic flux has zero classical energy,
Phys.\ Lett.\ B 97 (1980) 241;
%%CITATION = PHLTA,B97,241;%%
J. Groeneveld, J. Jurkiewicz and C.\,P. Korthals Altes,
%Twist as a probe for phase structure,
Phys.\ Scripta\ 23 (1981) 1022.

\bibitem{StaraLesna}
Ph.~de~Forcrand and O.~Jahn,
hep-lat/\discretionary{}{}{}0205026,
to be published in NATO ASI B.

\bibitem{PRL}
Ph.\ de Forcrand, M. D'Elia and M. Pepe,
%A study of the 't Hooft loop in SU(2) Yang-Mills theory,
Phys.\ Rev.\ Lett.\ 86 (2001) 1438.
%[hep-lat/0007034].
%%CITATION = HEP-LAT 0007034;%%

\bibitem{MMP}
A. Barresi, G. Burgio and M. M\"{u}ller-Preussker,
%Finite temperature phase transition, adjoint Polyakov loop and topology  in SU(2) LGT,
hep-lat/0209011, these proceedings.

\bibitem{G2}
K. Holland, P. Minkowski, U.-J. Wiese and M. Pepe, these proceedings.

\end{thebibliography}
\end{document}